\documentstyle[11pt,aaspp4,eqsecnum]{article}
\newfont{\myfont}{cmmib10}

\newcommand{\et}{{et al}\ }

\newcommand{\half}{{\textstyle{\frac{1}{2}}}}

\def\bg#1{\hbox{\boldmath$#1$}}
\newcommand{\bzeta}{{\bg\zeta}}

\newcommand{\bkappa}{{\bg\kappa}}

\begin{document}

\title{Scintillation-Induced Circular Polarization in Pulsars and Quasars}

\author{J.-P. Macquart$^{1}$ and D.B. Melrose$^{2}$\\
Research Centre for Theoretical Astrophysics\\
School of Physics, University of Sydney, NSW 2006, Australia.\\
$^{1}$jpm@physics.usyd.edu.au,
$^{2}$melrose@physics.usyd.edu.au
}

\date{\today}

\begin{abstract}
We present a physical interpretation for the generation of circular
polarization resulting from the propagation of radiation through a
magnetized plasma in terms of a rotation measure gradient, or `Faraday
wedges'.  Criteria for the observability of scintillation-induced circular
polarization are identified.   Application of the theory to the
circular polarization in pulsars and  compact extragalactic sources is
discussed.
\end{abstract}


\section{Introduction}

The circular polarization (CP) of radio emission observed from pulsars
(Manchester, Taylor \& Huguenin 1975; Radhakrishnan \& Rankin 1990; Han
\et 1999\markcite{Man75,RadRank90,Han98}) and from some quasars 
(Roberts \et 1975, Weiler \& de Pater 1983; Saikia \& Salter 
1988\markcite{JARob,Weiler,Saikia}) is not understood. In both
cases, simple theory suggests that any polarization should be linear,
determined by the direction of the projection of the magnetic field in
the source region on the plane of the sky. In both cases, intrinsic CP is
expected as a correction to first order in an expansion in the inverse
of the Lorentz factor of the radiating particles, but it does not seem
possible to account for the observations in terms of CP intrinsic to the
emission process (Radhakrishnan \& Rankin 1990; Radhakrishnan 1992;
Saikia \& Salter 1988\markcite{RadRank90,radhakrishnan,Saikia}). Another 
possibility is that the CP is imposed
as a propagation effect due to the partial conversion of linear
polarization into CP resulting from the ellipticity of the natural modes of the
medium (Sazonov 1969; Pacholczyk 1973; Jones \& O'Dell 
1977a,b\markcite{Sazonov,Pachol,Jones77a,Jones77b}). However,
this also fails to provide a satisfactory explanation for the properties
of the observed CP. In this paper we describe a different propagation
effect that can lead to CP for a source, independent of its degree of
linear polarization. We refer to this as scintillation-induced CP.
Although the observed CPs from pulsars and quasars are quite different,
we suggest that both might be due to scintillation-induced CP, with the
most obvious differences being due to pulsars scintillating in
the diffractive regime and quasars scintillating in the refractive regime.

In the formal theory of scattering in a turbulent, magnetized plasma
(Melrose 1993a,b\markcite{Melrose93a,Melrose93b}) there are differences in the scattering in the two
oppositely circularly polarized wave modes of the medium due to their
different refractive indices. Most of the terms contributing to the CP are
too small to be of interest for scattering in the interstellar medium
(ISM). However, we have recently shown (Macquart and Melrose 2000; 
hereinafter paper~I \markcite{MMcp1}) that there is one much larger contribution to the CP
that is a possible candidate for explaining the CP in pulsars and
quasars. The CP identified in paper~I is due to a nonzero variance in
Stokes $V$, and the formal theory implies that the dominant contribution
is of the form
$\langle V^2\rangle\sim D_{VV}(r_{\rm ref})\langle I\rangle^2$, where
$D_{VV}(r)$ is a phase structure function associated with the relative
phase, $\phi_V$, between the components in the opposite CPs. The distance
$r_{\rm ref}=r_{\rm F}^2/r_{\rm diff}$ is the refractive scale, which is
defined in terms of the Fresnel scale $r_{\rm F}=(L\lambda/2\pi)^{1/2}$,
where $L$ is the distance between the scattering screen and the
observer's plane and $\lambda$ is the wavelength, and the diffractive
scale $r_{\rm diff}$, which is defined by writing the phase structure
function in the forms $D(r)=(r/r_{\rm diff})^{\beta-2}$ for a power-law
spectrum of turbulence. Our purposes in this paper are threefold: first,
to provide a physical explanation for the mechanism that leads to this
term, second, to use this interpretation to relax some of the restrictive
assumptions made in paper~I to obtain a more general semiquantitative
expression for the predicted CP, and, third, to explore the suggested
application to the observed CP in pulsars and extragalactic sources.

We start (section~2) by including birefringence in a simple model for
strong scattering in which scattering is attributed to a large
number of coherent patches of size $\sim r_{\rm diff}$ within an envelope
of size $\sim r_{\rm ref}$ on the scattering screen (e.g., Goodman \& 
Narayan 1989, Narayan 
1992, Gwinn \et 1998\markcite{GoodNar,Narayan,Gwinnetal98}). When the birefringence is included, this model reproduces the
result
$\langle V^2\rangle\sim D_{VV}(r_{\rm ref})\langle I\rangle^2$
derived from the formal theory of scattering in a magnetized plasma in
paper~I. This simple model corresponds to strong diffractive
scintillation, suggesting that our expression for $\langle V^2\rangle$
applies only to a source that exhibits strong diffractive scintillations
(which is the case for pulsars but not for quasars). Further
physical interpretation of scintillation-induced CP is developed in
section~3, where it is argued that it arises from a combination of two
processes: a rippling of the wavefront, as in the conventional theory for
scattering in a turbulent medium, combined with random refractions in the
birefringent medium that cause a separation in the rays associated with the
opposite CPs. This leads to the following interpretation: the ripples in
the wavefront for the two CPs become spatially separated due
to the random birefringent refractions, so that they do not overlap in
the observer's plane. This leads to alternate patches in which one CP and
then the other dominates, leading to a nonzero $\langle V^2\rangle$. This
interpretation is the basis for two important generalizations that we
propose here. First, an essential requirement in this interpretation is
that the images in the two CPs be displaced relative to each other by
$\Delta x$ in the observer's plane. Our interpretation of the formal
theory is that this is due to random birefringent scatterings at the
putative scattering screen. However, any mechanism that causes such a
displacement of the rays corresponding to the two opposite CPs leads to a
nonzero $\Delta x$ and hence to a nonzero $\langle V^2\rangle$. In
particular, an alternative to random birefringent refractions is
birefringent refraction at a single structure, which we refer to as a
Faraday wedge, cf.\ section~4. Second, although the result derived from
the formal theory applies for strong diffractive scintillations, our
interpretation implies that a nonzero $\langle V^2\rangle$ results for
any source that exhibits scintillations. Consider a source with a
scintillation index $m_I$ on a spatial scale $r_{\rm scint}$ in the
observer's plane; it should exhibit fluctuations in the degree of CP,
$m_V$, that is of order $\Delta x/r_{\rm scint}$ times $m_I$. In
particular, a source that exhibits refractive scintillations should
exhibit fluctuations in CP on a similar timescale with an amplitude
smaller by the factor $\Delta x/r_{\rm scint}$.

For scintillation-induced CP to account for the observed CP in quasars and
pulsars, two conditions need to be satisfied. First, the predicted degree
of CP, $m_V$, must be in the observed range, which appears to be
$\sim1$ for a few pulsars and is $\ga10^{-3}$ for relevant extragalactic
sources. Second, the predicted timescale for the fluctuations in CP must
be consistent with the observed variations in the CP. For pulsars this is
the diffractive timescale, and for quasars it is the refractive timescale.
These requirements are discussed for pulsars in section~5, and for
extragalactic sources in section~6.

\section{Strong Scattering in a Birefringent Medium}
\label{strongscattering}

In this section we show that the main result found in paper~I is
reproduced by a simple model for the scattering. This result is
\begin{equation}
\langle V^2\rangle\sim D_{VV}(r_{\rm ref})\langle I\rangle^2,
\label{relation}
\end{equation}
which was derived in Paper I for scattering at a single screen for a
power-law spectrum of turbulence in a uniform magnetic field. The
assumption that the fluctuations are only in the density implies that
$D_{VV}(r)$ is proportional to the phase structure function, $D(r)$, for
the phase in the absence of birefringence. Assuming a power-law
distribution for the turbulence, this implies
\begin{equation}
D_{VV}(r)=\alpha^2\left(\frac{r}{r_{\rm diff}}\right)^{\beta-2},
\label{Dr}
\end{equation}
with $\beta=11/3$ for a Kolmogorov spectrum
of turbulence, and with $\alpha=Y\cos\theta$, $Y=\nu_B/\nu$, where 
$\nu_B$ is the electron cyclotron frequency and $\theta$ is the angle 
between the magnetic field and line of sight.

The model for strong scattering that we introduce to interpret the result
(\ref{relation}) involves regarding the amplitude of the wave at the observer's screen as
a sum of terms that each correspond to a point of stationary phase (e.g.,
Born \& Wolf 1965, Goodman 1985, Gwinn \et 
1998\markcite{BW,Goodman85,Gwinnetal98}). The model is illustrated 
in Figure~1. There are two
relevant contributions to the phase: a rippling on a scale $r_{\rm diff}$
due to the postulated density fluctuations at the screen, and a geometric
effect described by a curvature of the mean wavefront on a scale $r_{\rm
F}$, with the points of stationary phase corresponding to points where
the tangent to the wavefront is orthogonal to the line of sight, cf.\
Figure~1. Let there be
$N\sim(r_{\rm F}/r_{\rm diff})^4\gg1$ such points of stationary 
phase (e.g. \markcite{Narayan}Narayan 1992). The
amplitude in this model is of the form
\begin{equation}
u=\sum_{j=1}^NA_je^{i\phi_j},
\label{phasors}
\end{equation}
where the $A_j$ and $\phi_j$ are the amplitude and phase, respectively,
associated with the $j$th coherent patch (the $j$th point of stationary
phase). The $\phi_j$ are assumed to be a random set of phases uniformly
distributed in the interval $[0,2\pi]$. For our purposes it suffices to
assume that all $N$ amplitudes are identical, writing $A_j=A$ for all
$j$. The mean is defined by averaging over the phases. The intensity is
\begin{equation}
\langle I\rangle=\langle uu^*\rangle=
\left\langle\sum_{j,k=1}^NA^2e^{i(\phi_j-\phi_k)}\right\rangle
=NA^2,
\label{intensity}
\end{equation}
and its variance is
\begin{equation}
\langle I^2\rangle=\langle uuu^*u^*\rangle=
\left\langle\sum_{j,k,l,m=1}^NA^4
e^{i(\phi_j+\phi_k-\phi_l-\phi_m)}\right\rangle
=N(2N-1)A^4,
\label{varianceI}
\end{equation}
For $N\gg1$ these imply $\langle I^2\rangle=2\langle I\rangle^2$, which
is the well-known result for diffractive scintillations.

To include the polarization the wave amplitude, $u$, is separated into its
right-hand, $u_+$, and left-hand, $u_-$, circularly polarized components,
and the intensities in the two CPs are identified as
$I_\sigma=u_\sigma u_\sigma^*$ with $\sigma=\pm$. The relevant Stokes
parameters are given by
\begin{equation}
I=\frac{1}{2}(I_++I_-),
\qquad
V=\frac{1}{2}(I_+-I_-).
\end{equation}
The difference in refractive
index between wave in the two CPs leads to a phase difference between
them, denoted $\sigma\phi_V$. Let $\sigma\phi_{Vj}$ be the
additional phase associated with the $j$th coherent patch.
This assumption corresponds to
\begin{equation}
u=\sum_{\sigma=\pm}u_\sigma,
\qquad
u_\sigma=\sum_{j=1}^NA_je^{i(\phi_j+\sigma\phi_{Vj})},
\label{phasors2}
\end{equation}
The mean CP is then
\begin{equation}
\langle V \rangle=A^2
\left\langle\frac{1}{2}\sum_{i,j=1}^N
\left[e^{i(\phi_i+\phi_{Vi}-\phi_j-\phi_{Vj})}
-e^{i(\phi_i-\phi_{Vi}-\phi_j+\phi_{Vj})}\right]\right\rangle.
\label{aveV1}
\end{equation}
We assume that the change in phase due to the birefringence at the screen
is sufficiently small that the approximation
\begin{equation}
e^{i(\phi_j+\sigma\phi_{Vj})}=
(1+i\sigma\phi_{Vj}-\textstyle{{1\over2}}\phi_{Vj}^2)e^{i\phi_j}
\label{approx1}
\end{equation}
applies. On substituting (\ref{approx1}) into (\ref{aveV1}), an average
over the random phases is nonzero only for $i=j$ and the two terms in
(\ref{aveV1}) cancel for $i=j$, giving $\langle V\rangle=0$.

The variance in $V$ is given by
\begin{equation}
\langle V^2 \rangle =\langle I_+^2 + I_-^2 - 2 I_+ I_- \rangle,
\label{aveV2}
\end{equation}
which becomes
\begin{equation}
\langle V^2 \rangle=
\frac{A^4}{4}
\sum_{i,j,k,l=1}^N
\left\langle
e^{i(\phi_i+\phi_j-\phi_k-\phi_l)}
\bigg[\sum_{\sigma=\pm}
e^{i\sigma(\phi_{Vi}+\phi_{Vj}-\phi_{Vk}-\phi_{Vl})}
-2e^{i(\phi_{Vi}-\phi_{Vj}-\phi_{Vk}+\phi_{Vl})}
\bigg]\right\rangle,
\label{aveV3}
\end{equation}
On expanding as in (\ref{approx1}), the only nonvanishing terms are for
$i=l$ and $j=k$. There are $N^2$ such terms, which give
\begin{equation}
\langle V^2 \rangle=N^2A^4D_{VV}(r_{ij}),
\qquad
D_{VV}(r_{ij})=\langle(\phi_{Vi}-\phi_{Vj})^2\rangle,
\label{aveV4}
\end{equation}
where $D_{VV}(r_{ij})$ is the phase structure function for $\phi_V$. The
distance $r_{ij}$ is the spatial separation of the $i$th and $j$th
coherent patches, which is typically of order
$r_{\rm ref}$. With  $NA^2=\langle I\rangle$ according to
(\ref{intensity}), the result (\ref{aveV4}) with $r_{ij}\to r_{\rm ref}$
reproduces the functional form of the relation (\ref{relation}).

The numerical coefficient in the relation (\ref{relation}) can be
evaluated for a power-law spectrum of the turbulence, as given by
(\ref{Dr}). A calculation given in Appendix~A leads to a coefficient,
$g(\beta)$, of order  unity. Thus (\ref{aveV4}) with (\ref{intensity})
implies
\begin{equation}
\langle V^2 \rangle=g(\beta)D_{VV}(r_{\rm ref})\langle I\rangle^2,
\label{aveV5}
\end{equation}
with $D_{VV}(r)$ given by (\ref{Dr}).

The expressions (\ref{aveV4}) or (\ref{aveV5}) are valid only for
$\langle V^2 \rangle\ll\langle I\rangle^2$; this follows from the
expansion made in (\ref{approx1}) and a related expansion for
$D_{VV}(r)\ll1$ made in paper~I. However, there is no reason in principle
why this condition must be satisfied. If it is invalid, then the
fluctuations in $V$ can approach the maximum possible value,
$\langle V^2 \rangle\sim\langle I\rangle^2$.

This simple model for $\langle V^2\rangle$ suggests the following
interpretation for the result (\ref{relation}). The assumptions made in
the calculation in paper~I apply to a source that, in the absence of
birefringence, exhibits strong diffractive scintillations. The inclusion
of birefringence leads to random birefringent refractions that cause the
two oppositely circularly polarized rays to emerge from the scattering
screen propagating in slightly different directions. This angular
separation between the rays implies that the peaks and troughs in the
wavefronts in the opposite CPs are displaced from each other in the
observer's plane. This separation corresponds to alternating patches of
opposite CPs, which is described by $\langle V^2\rangle$.

\section{Random Birefringent Refractions}
\label{RBR}

In this section we discuss the interpretation of the function
$D_{VV}(r_{\rm ref})$ in (\ref{relation}) in terms of random birefringent
refractions. We then argue that this interpretation allows one to relax
two restrictive assumptions made in paper~I to obtain a more general,
semiquantitative expression for the predicted scintillation-induced CP.

Random refractions in an isotropic medium lead to an angular (`scatter')
broadening of the source. In scattering theory this is described in terms
of a bundle of rays diffusing in angle of propagation. For our
purposes here a semiquantitative discussion suffices. In refractive
scattering, the screen acts like a large lens, of size
$\sim r_{\rm ref}$, that causes a phase change of order
$\delta\phi\sim [D(r_{\rm ref})]^{1/2}$. The phase change includes a tilt
of the wavefront that corresponds to a change in the ray direction. The
typical change in the ray angle is
$\delta\psi\sim(\lambda/2\pi)\delta\phi/r_{\rm ref}$. When the
birefringence is included, there is an additional phase change between
the two wave modes, and this leads to a separation of the ray angles for
the opposite CPs. The relative phase change between the wavefronts for
the two CPs is of order $\delta\phi_V\sim [D_{VV}(r_{\rm ref})]^{1/2}$,
and the angular separation between the rays in the two CPs is
$\Delta\psi_V\sim(\lambda/2\pi)\delta\phi_V/r_{\rm ref}$.
This angular separation times the distance, $L$, to the screen leads to a
spatial displacement of the images in the two CPs of
$\Delta x\sim\Delta\psi_VL$, which implies
\begin{equation}
D_{VV}(r_{\rm ref})\sim\frac{(\Delta x)^2}{r_{\rm diff}^2}.
\label{DVV}
\end{equation}
The expression for $\langle V^2 \rangle$ obtained in paper~I, and
rederived in (\ref{aveV5}) above, then implies
 \begin{equation}
\langle V^2 \rangle\sim\frac{(\Delta x)^2}{r_{\rm diff}^2}
\langle I\rangle^2.
\label{aveV6}
\end{equation}

% Before discussing the interpretation of (\ref{aveV6}), a preliminary
% point concerns scintillations in the absence of birefringence. Then the
% model in section~2 leads to diffractive scintillations in the total
% intensity, corresponding to a scintillation index $m_I\sim1$. The
% interpretation is that there are patches of constructive and destructive
% inference in the observer's plane, of typical size
% $r_{\rm scint}\sim r_{\rm diff}$ for diffractive scintillations. We refer
% to this in terms of ripples in the wavefront having a characteristic size
% $\sim r_{\rm scint}$.

We interpret the form (\ref{aveV6}) as follows. According to the model in
section~2, the source is exhibiting diffractive scintillations that
correspond to a rippling of the wavefront such that there are patches of
constructive and destructive interference with a characteristic size
$r_{\rm diff}$. The birefringence causes a relative spatial displacement
by $\Delta x$ of this pattern in the opposite CPs. For
$\Delta x\la r_{\rm diff}$, the partial separation of the images in the
opposite CPs leads to an image with a degree of CP
$\sim\Delta x/r_{\rm diff}$.

The foregoing discussion shows that two ingredients are essential for
scintillation-induced CP: scintillations that
produce a pattern of variations with a scale size $r_{\rm scint}$ in the
observer's plane, and birefringent refraction that causes a displacement,
$\Delta x$, between the images in the opposite CPs. The resulting degree
of CP is then given by
\begin{equation}
m_V\sim\frac{\Delta x}{r_{\rm scint}}\,m_I,
\qquad
m_V=\frac{\langle V^2\rangle^{1/2}}{\langle I\rangle},
\qquad
m_I=\frac{[\langle(I^2\rangle-\langle I\rangle^2]^{1/2}}
{\langle I\rangle}.
\label{aveV7}
\end{equation}
With the assumptions made in the models discussed above, the
scintillations are diffractive, $m_I\sim1$,
$r_{\rm scint}=r_{\rm diff}$, and $\Delta x$ is due to random
birefringent refractions. More generally, (\ref{aveV6}) should apply with
$m_I$, $r_{\rm scint}$ corresponding to the observed scintillations, and
$\Delta x$ could be due to birefringent refraction either by a single
structure (Faraday wedge) or by a spectrum of turbulence.  For
example, for refractive scintillations one has
$m_V=(\Delta x/r_{\rm ref})m_{\rm ref}$, where $m_{\rm ref}$ is the
intensity modulation due to  refractive scintillation.

\section{The Faraday Wedge}\label{FaradayWedge}
The random birefringent refractions that contribute to
scintillation-induced CP through the factor $D_{VV}(r_{\rm ref})$ in
(\ref{relation}) are due to gradients in rotation measure (RM) 
associated with the turbulence at the scattering screen.  Scintillation-induced CP requires
that the wavefront be rippled and that the ripples in the opposite CPs be
displaced laterally along the wavefront, but it is not necessary that
these two features be imposed at a single scattering screen. In this
section we introduce the concept of a Faraday wedge in which the lateral
displacement along the wavefront of the ripples in the opposite CPs is
attributed to birefringent refraction at a single structure along the
line of sight.  We consider two idealized models for the wedge; in the
first the wedge is modeled as a birefringent region with a density
gradient and in the second the wedge is modeled as a prism with sharp
edges. These two models lead to essentially the same semiquantitative
result for the angular separation of the rays corresponding to the two
CPs.

\subsection{Refraction in a Birefringent Medium}\label{Birefringent}

It is useful to view the effects of a Faraday wedge in terms of both
geometric optics (rays) and physical optics (wavefronts). In terms of
geometric optics, as illustrated in Figure~2, a Faraday wedge splits an
incident ray into two oppositely circularly polarized emerging rays
propagating in slightly different directions, separated by an angle
$\Delta\,\zeta$ say. After propagating the distance $L$ from the wedge to
the  observer's plane, the left- and right-hand images are separated by
$\Delta x=\Delta\,\zeta\,L$.  For $\Delta\,\zeta\,L\gtrsim r_{\rm scint}$ 
an observer sees
well-separated oppositely circularly polarized images of the source for a
scintillating source. In terms of
physical optics, as illustrated in Figure~3, the Faraday wedge causes an
incident wavefront to split into two wavefronts whose normals are in
slightly different directions, separated by $\Delta\,\zeta$. Each
wavefront is rippled, and the effect of the Faraday wedge applies to the
average (over the ripples) wavefronts. (One wavefront is also slightly delayed
relative to the other, but this effect is neglected here.) At the
observer's plane, corresponding ripples on the two wavefronts are
displaced along the wavefront by $\Delta x= \Delta\,\zeta\,L$. The observer sees a
net CP varying on the timescale associated with the scintillations as
the whole pattern sweeps across the observer's plane.

The following estimate of the angle $\Delta\,\zeta$ applies when there is
a smooth gradient in RM. Consider a planar wavefront propagating parallel
to the $z$-axis incident upon a slab of material of thickness $\Delta z$,
containing spatial variations in the plane transverse to the $z$-axis. As
shown in Appendix~A, using the paraxial approximation, the  angular
deviation of the ray in terms of the refractive index variations on the
scattering screen is determined by
\begin{eqnarray}
{d(n_\sigma\,{\bkappa})\over dz}=
{\partial n_\sigma\over\partial{\bf r}},
\end{eqnarray}
where $\sigma=\pm$ denotes the two wave modes, ${\bf r}=(x,y)$ is the plane orthogonal to the $z$-axis,
${\bkappa}={\bf k}/k$ is the direction of the wavevector and the
arguments of the refractive index $n_\sigma(\omega,{\bkappa},z,{\bf r})$
are suppressed. In the weak anisotropy limit one writes the refractive
index in  terms of polarization dependent and independent terms
$n_\sigma=n+\sigma\Delta n/2$ with $\Delta n=n_+-n_-$.
For a screen a distance $L$ from the observer, the
relative displacement of the centers of the images in the two
modes on the observer's plane is $L\Delta {\bzeta}$ with
\begin{eqnarray}
\Delta {\bzeta}=-{\Delta n\over n}\frac{\partial}{\partial{\bf r}}
\int_0^{\Delta z}dz\,n
+\frac{\partial}{\partial{\bf r}}\int_0^{\Delta z}dz\,
\Delta n.
\label{deltazeta1}
\end{eqnarray}
The refractive indices depend on the magnetoionic
parameters, $X=(\nu_p/\nu)^2$, $Y=\nu_B/\nu$, where $\nu_p$ is the 
plasma frequency and $\nu_B$ is the electron cyclotron frequency.  
One has $X\ll1$, $Y\ll1$ here, with $n=1-\half X$,
$\Delta n=XY\cos\theta$, where $\theta$ is the angle between the
wave-normal direction and $\bf B$. Then dominant contribution to
$\Delta{\bzeta}$ is due to the gradient of
$\int_0^{\Delta z}dz\,XY\cos\theta=\Delta{\rm RM}\,\lambda^3/2\pi$, where
$\Delta{\rm RM}$ is the contribution of the path length $\Delta z$ to the
RM. We also introduce the RM phase,
$\phi_{\rm RM}=\lambda^2\,{\rm RM}$, which is the relative phase
difference between the two wavefronts.  The displacement  between the
left- and right-circularly polarized  images on the observer's screen
reduces to (further details are given in Appendix~B)
\begin{eqnarray}
{\bf\Delta x}({\bf r})=\frac{L\lambda}{2\pi}\,
\nabla_{\perp}\phi_{\rm RM},
\label{WedgeDisp}
\end{eqnarray}
where $\nabla_\perp$ denotes the gradient in the $x$-$y$ plane.

\subsection{Refraction at a Faraday prism}
\label{Faradayprism}
An alternative model for a Faraday wedge is a prism. Let the prism have
an apex angle of $\psi$ and let its refractive index be $n-n_0$
greater than the refractive index ($n_0$) in the surroundings. For
simplicity, suppose that the prism is oriented nearly perpendicular to
the line of sight, so that we are interested only rays at small angles
relative to the direction perpendicular to the axis of the prism, cf.\
Figure~3. Then a ray incident at an angle $\theta_{\rm in}$ emerges at an
angle
$\theta_{\rm out}$ given by
\begin{equation}
\theta_{\rm out}=\theta_{\rm in}-2(n-n_0)\,\tan(\psi/2).
\label{inout}
\end{equation}
The mean angular deviation,
$\Delta\theta=\theta_{\rm out}-\theta_{\rm in}$, of a ray at the prism
and the difference, $\Delta\,\zeta$, between the emerging ray in the two
CPs are then given by
\begin{equation}
\Delta\theta=2(n-n_0)\,\tan(\psi/2),
\qquad
\Delta\,\zeta
=2\Delta n\,\tan(\psi/2),
\label{deltazeta2}
\end{equation}
respectively.

Let us compare the results (\ref{deltazeta1}) and (\ref{deltazeta2}).
Retaining only the final term in (\ref{deltazeta1}), assuming the region
to be uniform along the $z$ direction and with a uniform gradient with a
scale length $L_\perp$ in the perpendicular direction,
(\ref{deltazeta1}) gives $\delta\,\zeta=\Delta n\,\Delta z/L_\perp$. Thus
the two results coincide for $2\tan(\psi/2)=\Delta z/L_\perp$. More
generally one can conclude that a Faraday wedge leads to a deviation of
the rays with opposite CPs by an angle $\Delta\,\zeta$ which is of order
the difference, $\Delta n$, in the refractive indices of the two modes
within the wedge, times a geometric factor ($\Delta z/L_\perp$ or
$2\tan(\psi/2)$) which is less than or of order unity.

\subsection{Requirements on the Model}\label{Requirements}
A Faraday wedge leads to CP due to a displacement, $\Delta x$, 
between ripples in the wavefronts corresponding to opposite CPs.  Suppose 
that the ripples have a scale length $r_{\rm scint}$ and that the
modulation index, $m_I$, of the intensity due to these ripples. Then
one should observe fluctuations in CP with an rms degree of CP of
\begin{equation}
\frac{V_{\rm rms}}{I}\sim m_I\cases{
\Delta x/r_{\rm scint}&for $\Delta x\ll r_{\rm scint}$,
\cr
1&for $\Delta x\ga r_{\rm scint}$,
\cr}
\label{CP1}
\end{equation}
on a characteristic timescale $r_{\rm scint}/v$, where $v$ is the speed at
which the pattern moves across the observer's plane.  For diffractive
scattering one has $r_{\rm scint}\sim r_{\rm diff}$ and $m_I\sim1$, and for
refractive scattering one has $r_{\rm scint}\sim r_{\rm ref}$ and
$m_I\sim(r_{\rm diff}/r_{\rm ref})^{(\beta-4)/2}$, with
$(\beta-4)/2=-1/6$ for a Kolmogorov spectrum $\beta=11/3$.
For a source of angular size $\theta_s$ to exhibit scintillations
requires $\theta<r_{\rm diff}/L$ and $\theta<r_{\rm ref}/L$
for diffractive and refractive scattering, respectively.

A semi-quantitative estimate for when these conditions are satisfied
is deduced from equation (\ref{aveV7}) as follows. With
$\phi_{\rm RM}={\rm RM}\lambda^2$, we require
$\Delta x\ga r_{\rm diff}$ for diffractive scintillations and
$\Delta x\ga r_{\rm ref}$ for refractive scintillations. Hence we require
\begin{equation}
\frac{\nabla_\perp{\rm RM}\lambda^3}{2\pi}
\ga\cases{
\displaystyle{
\frac{r_{\rm diff}}{L}}\ga\theta_s
&\quad\hbox{diffractive},
\cr\noalign{\vskip3pt}
\displaystyle{
\frac{r_{\rm ref}}{L}}\ga\theta_s
&\quad\hbox{refractive}.
\cr}
\label{condition2}
\end{equation}
If the angular deviation, $\Delta\,\zeta$, between the rays in the two
modes is dominated by a single structure (single Faraday wedge), which
contributes $\Delta\,$RM to the total RM, then one has
$\nabla_\perp\phi_{\rm RM}\sim\Delta{\rm RM}\lambda^2/L_\perp$, where
$L_\perp$ is a distance that characterizes the gradient in RM.
The condition (\ref{condition2}) includes the requirement
$\Delta{\rm RM}\lambda^2/2\pi L_\perp\ga\theta_s$. Assuming that the
Faraday wedge has a density $n_e$ and a magnetic induction $B$,
$\Delta\,$RM is proportional to $n_eB$ times the line of sight distance,
$\Delta z$ through the structure.
% Provided that $n_e$ is substantially
% different from the surrounding density along the line of sight distance
% through the structure.
Then the condition (\ref{condition2}) reduces to
${\rm RM}\lambda^3 /2\pi L_\perp\ga\theta_s$, and on inserting numerical
values, this becomes
\begin{equation}
\lambda^3\,\left(\frac{n_e}{{1\rm\,m}^{-3}}
\right)\,\left(\frac{B}{{1\rm\,G}}\right)
\left(\frac{\Delta z}{L_\perp}\right)
\ga10^9\left(\frac{\theta_s}{{1\rm\,mas}}\right).
\label{condition3}
\end{equation}
Physically, the dependence on $\Delta z/L_\perp$ is due to the
requirement that refraction cause a large enough separation between the
rays in the two CPs. The region acts like a prism with a small angle at
its apex; the deviation of the ray is zero when this angle is zero (the
prism reduces to a slab), and the deviation increases as this angle
increases.

The condition (\ref{condition3}) is not easily satisfied: it requires an
exceptionally dense, strongly magnetized region along the
line of sight, and it is more easily satisfied at longer wavelengths.

\section{Application to Pulsars}
The degree of CP observed in most pulsars is relatively small
(e.g., Han \et 1999\markcite{Han98}). However, the degree of CP usually
quoted refers to the pulse-averaged quantity. Studies of single pulses are
possible for a small subset of pulsars, and the CP in single pulses can
be very much greater than the pulse-averaged CP
(Manchester, Taylor \& Huguenin 1975\markcite{Man75}). Thus the
pulse-averaged CP appears to be the mean value of a CP that can vary
greatly on a timescale of order the pulse period.  There is no
satisfactory explanation for the pulse-averaged CP (e.g., Radhakrishnan
\& Rankin 1990; Kazbegi, Machabeli \& Melikidze 1991; Radhakrishnan 
1992\markcite{RadRank90,Kazbegi,radhakrishnan}).
It is usually assumed this is imposed either by the emission process 
or by propagation in the pulsar magnetosphere.  In this section we discuss
the possibility that this rapidly varying component in the CP might be
due to the Faraday wedge effect associated with diffractive
scintillations induced by propagation through the ISM.

\subsection{The Vela pulsar}\label{Vela}
We apply the foregoing ideas to the Vela pulsar for which there is
direct evidence for RM fluctuations.
Three parameters are required to estimate the magnitude of the CP generated
by the Faraday wedge: the magnitude of the RM gradient,
the slope of the power spectrum of density inhomogeneities, $\beta$, and
the refractive scale, $r_{\rm ref}$.  We list our best
estimates of these parameters, and then use these to estimate the
degree of CP expected according to the foregoing
theory.

Hamilton, Hall \& Costa (1985)\markcite{Hamiltonetal} reported a linear
change in RM across the Vela pulsar for the interval
1970-1985.  The best fit to the RM gradient is 0.73
rad/m$^2$/yr, which translates into a spatial RM gradient of
$2.3 \times 10^{-11} v_{\rm km/s}^{-1}$ rad$\,$m$^{-3}$, where $v_{\rm
km/s}$ is
the speed of the wedge
transverse to the line of sight in km$\,$s$^{-1}$.
We assume the Kolmogorov value of $\beta=11/3$ for the power spectrum of
density inhomogeneities for this pulsar.  This is consistent with
some measurements, although we note that the measurements of
refractive flux variations are more consistent with $\beta$ closer to
3.9 (Johnston, Nicastro \& Koribalski 1998\markcite{Johnstonetal}).
The size of the scattering disk, as measured by \markcite{Gwinnetal}Gwinn
(1997), is $\approx 1.0$ AU at around 2.3 GHz.  (The actual measurement band
was from 2.273 to 2.801 GHz).
The refractive length, $r_{\rm ref}$, scales as
$\nu^{-[1+2/(\beta-2)]}$.

In estimating the Faraday rotation phase change across the scattering
disk, we choose an observing frequency of 600 MHz, where the effect of
Faraday rotation is strong.  Scaling the refractive length to this
frequency, one has $r_{\rm ref} \approx 19\,$AU.  Combining this with our
estimate of the RM gradient, the total RM change across the scattering
disk, $r_{\rm ref} \nabla_\perp \phi_{\rm RM}$, is $15.9 v_{\rm
km/s}^{-1}$. To calculate the root-mean-square degree of CP we use this
estimate of $r_{\rm ref}$ in equation (\ref{aveV7}) for diffractive
scintillation, giving $\sqrt{\langle V^2\rangle(z)/
\langle I^2 \rangle (0)} = 15\% \,(4\%)$, assuming that
$v_{\rm km/s}$ is 100 (500).

We conclude that measurement of scintillation-induced
CP is feasible for the Vela pulsar, and that it is
likely to be feasible for other pulsars at low frequency.
Although not discussed in detail here, there are several other pulsars
known to exhibit variability in RM, including the Crab pulsar (Rankin et
al.\ 1988\markcite{Rankin}) and PSR~1259-63 (Johnston \et 1986\markcite{Johnston1259} ), which exhibits RM variations of $\sim 200$ rad\,m$^{-2}$ on
timescales of 0.5 hr. An observational test of the scintillation-induced
CP model requires statistics on the variance in the CP in individual
pulses, and the way this variance changes with frequency of observation.

\section{Application to Extragalactic Sources}
A small but significant degree of CP ($\lesssim 0.1$\% to a few \%)
is observed in some compact extragalactic radio sources
(e.g., Roberts \et 1975, Weiler \& de Pater 1982, de Pater \& Weiler 1982,
Komesaroff \et 1984\markcite{JARob,Weiler,dePater,Kom}).  The suggested
interpretations include the intrinsic polarization associated with
synchrotron radiation (Legg \& Westfold 1968\markcite{LeggWest}), and
partial conversion of linear into
circular polarization due to ellipticity of the natural wave modes of
the cold background plasma (Pacholczyk 1973\markcite{Pachol}) or of the
relativistic electron gas itself (e.g., Sazonov 1969,
Jones \& O'Dell 1977a,b\markcite{Sazonov,Jones77a,Jones77b}).  However 
none of these
suggested interpretations has proved satisfactory in accounting for
(a) the frequency dependence, (b) the temporal variations, and (c)
the magnitude of the observed CP.  Here we explore the possibility that
the CP observed in compact extragalactic sources is scintillation-induced
CP. We consider the requirements for scintillations to produce 0.1\% CP
due to scintillation either in our Galaxy or in the host object.

\subsection{Extragalactic sources}\label{ExtraEst}

Scintillation-induced CP relies on birefringent refraction, which is
determined by the gradient in RM.  To estimate the degree of CP using 
equation (\ref{aveV7}) requires an
estimate of the parameter $\Delta x$, which is determined by
(\ref{DVV}) with (\ref{Dr}). Our estimate for the CP then reduces to
\begin{eqnarray}
    \sqrt{\frac{\langle V^2 \rangle (z)}{\langle I^2 \rangle(0)} }
\sim \alpha
    \left( \frac{ r_{\rm F} }{ r_{\rm diff} }\right)^{\beta-2}.
\label{ISMV}
\end{eqnarray}
where we ignore an observed slight excess in the relative
fluctuations in RM in our Galaxy compared with the relative fluctuations
in the electron density (Minter \& Spangler 1996\markcite{MinSpa}).

We estimate the degree of CP at 5~GHz.
For a screen at distance of $L=1\,$kpc, the Fresnel scale is
$r_{\rm F} = \sqrt{\lambda L/2 \pi} = 5 \times 10^8\,$m.  The length
scale $r_{\rm diff}$ is estimated from the frequency,
$\nu_t$, at which the scattering becomes strong, at $r_{\rm diff} =
r_{\rm F}$, and from the fact that $r_{\rm diff}$ scales proportional
to $\nu^{2/(\beta-2)}$.  We assume a Kolmogorov spectrum of density
inhomogeneities, $\beta=11/3$.  For an object located off the Galactic
plane one typically has $\nu_t \approx 7\,$GHz (Walker 
1998\markcite{Walker}),
yielding $r_{\rm diff} \approx 3 \times 10^8\,$m at 5~GHz.
Taking $\langle B_G \cos \theta \rangle = 3 \,\mu$G one has
$\alpha\approx 2.5 \times 10^{-9}$ which yields
$V_{\rm rms} = 6 \times 10^{-9} \langle I^2 \rangle^{1/2}$.
Hence, we conclude that the contribution of Galactic RM fluctuations to
the CP is negligible.

Next consider RM variations internal to an extragalactic source. The
theory (developed in Paper I) needs to be modified slightly if the
Faraday wedge is assumed to be located in the host object because the
non-planarity of the wavefront cannot be ignored. The modifications for
a spherical wavefront involve making the substitutions (Goodman \& 
Narayan 1989\markcite{GoodNar})
\begin{eqnarray}
    z \rightarrow \frac{z_1 z_2}{z_1 + z_2}, \label{Corre1}\\
    {\bf r} \rightarrow \frac{z_1}{z_1 + z_2}{\bf r} \label{Corre2},
\end{eqnarray}
where $z_1$ is the distance from the source to the scattering screen and
$z_2$ the distance from the screen to the observer. Using equations (4.9)
in (5.21) in Paper~I, this implies
$\langle V^2 \rangle^{1/2} \approx 2^{-1/2} (z_1 \nabla_{\perp} \lambda^2
{\rm RM}/(2 \pi r_{\rm diff}/\lambda))^{(\beta-2)/2}
\langle I^2 \rangle^{1/2}$.

For the sake of discussion we choose the
specific object 3C~345 for which there is an estimated dependence
(Matveenko \et 1996\markcite{Matveenko}) of ${\rm RM} = 3500 (R/3.79h^{-1} \, {\rm
pc})^{-3}\,$rad/m$^2$ on radial distance $R$ from the core, where
$R=3.79h^{-1}\,$pc corresponds to an angular size of 1~mas.
Consider the RM gradient required to produce 0.1\% CP at 5~GHz.
Since the scattering is likely to be concentrated around the AGN core we
take an effective distance of $z_1=100\,$pc between the source and
screen and $r_{\rm diff} = 10^7\,$m.  One then requires
$\nabla_\perp {\rm RM} \approx 3.6 \times 10^{-11}$ rad/m$^3$.
We conclude that variations in the large-scale RM are
large enough to produce this degree of CP within
the central $0.85 \, h^{-3/4}\,$pc of the core.

The distance between the source and the screen is assumed to be $z_1 = 
100$~pc.  There is no direct evidence for this parameter and our 
choice is based on a plausibility argument in view of the known 
scattering parameters of Sgr~A*.  
Sgr~A* is an AGN-like object at our own Galactic center which appears to 
exhibit refractive scintillation (e.g., Zhao \& Goss 1993\markcite{zhao93}), 
presumably due to turbulence in a medium $\sim 100$~pc from the 
source (\markcite{Backer78}Backer 1978).  Our plausibility argument is 
simply that compact extragalactic sources are likely to have some 
similarities to Sgr~A*, and that it is plausible that they 
scintillate  due to turbulence in the surrounding interstellar 
medium $\sim 100$~pc from the source.

The foregoing estimates apply when the source exhibits diffractive
scintillation.  CP can also result from refractive scintillation,
for which the source size requirement is less stringent.  However, the
requirement on the separation of the two senses of CP is more stringent:
the Faraday wedge must cause a separation that is a significant fraction
of the scale on which refractive flux variations occur. Thus, one
requires (cf.\ equation (\ref{aveV7})) $\langle V^2 \rangle^{1/2}
\approx m_{\rm ref} \, r_{\rm diff} \lambda^2 \nabla_\perp {\rm RM},$
where $m_{\rm ref}\approx (r_{\rm diff}/r_{\rm F})^{2-\beta/2}$ is the
refractive modulation index. For example, the RM gradient in 3C~345 used
above is sufficiently large that a centrally-located object small enough
to exhibit refractive scintillation would produce $>0.1\%$ CP at 5~GHz
provided the RM gradient is greater than $3 \times 10^{-8}\,$rad/m$^3$.
According to the RM model used above, such gradients are encountered
within the central $0.15 \,h^{-3/4}\,$pc of the core.

Similar data available on large RM gradients on milliarcsecond scales near
the cores of other extragalactic objects (\markcite{DHRob}Roberts \et
1990, \markcite{Udom}Udomprasert \et 1997, \markcite{Caw}Cawthorne et
al.\ 1997, \markcite{Taylor}Taylor 1998) suggests that
CP due to this effect is possible for a broad class of extragalactic
sources.

The timescale of variability in the CP depends on the relative 
speed, $v_s$, of the screen and the
source and on the relative speed, $v_o$ of the observer and the screen.
The speed at which the pattern moves across the observer's plane is $v_s
(z_1+z_2)/z_1$, and the observer moves across this plane at $v_o$.  
The pattern scale at the observer is $r_{\rm diff} (z_1+z_2)/z_1$ for 
diffractive scintillation and $r_{\rm ref} (z_1+z_2)/z_1$ for refractive 
scintillation.  The
timescale for variations in $V$ is determined by the pattern scale of 
the oppositely circularly polarized regions divided by the relevant
speed.  For a point source, diffractive scintillation would be observed
on a timescale $r_{\rm diff}/v_s$, and refractive scintillation would be
observed on a timescale $r_{\rm ref}/v_s$, where it is assumed that
the relevant speed is that of the screen relative to the source.  
This is the case for $z_1 \ll z_2$, since the apparent speed of 
the scattering material across the line of sight dominates 
the motion of the observer across the pattern.

Taking $v_s=100\,$km/s with the parameters assumed above, the timescale
of diffractive variations is $10^2\,$s and the refractive
timescale is $3 \times 10^4\,$s.  Based on these estimates, diffractive
variations should exhibit reversals in sign of the CP on a timescale of a
few minutes, which does not explain the observed variability over hours
to days (\markcite{Kom}Komesaroff \et 1984). Although the refractive timescale is closer to observed timescale
of CP variations in extragalactic sources, evidence for
reversals in the sense of the CP on this timescale is lacking.

\section{Conclusions}
It was shown in Paper I that propagation of radio waves through a
magnetized  inhomogeneous plasma leads to fluctuations in  
circular polarization (CP), even if the scintillating  source is 
itself unpolarized.  Although the mean CP
induced by propagation through the medium is zero, the variance,
$\langle V^2 \rangle$, is nonzero.  

The generation of this CP is attributed to a gradient in RM, called a
Faraday wedge, leading to a lateral displacement at the observer's plane
of the wavefronts for opposite CPs. Inhomogeneities in the ISM introduce
corrugations into the wavefront, and the displacement of these at the
observer's plane leads to alternate regions of opposite CP. As
the ISM moves across the line of sight to a source, an observer samples
these alternate patches of opposite CPs. The mean CP is zero, and the
variance is nonzero. The effect is significant provided that the lateral
displacement of the corrugations in the wavefronts for the opposite CPs
is a significant fraction of (or larger than) the typical size of the
corrugations. This scintillation-induced CP varies on the timescale
associated with the scintillations.  For a source undergoing diffractive
scintillations, the CP varies on the diffractive timescale, and for a
source undergoing refractive scintillations, the CP varies on the
refractive timescale.

Pulsars exhibit both diffractive and refractive scintillations, and
one expects them to exhibit scintillation-induced CP on a timescale
associated with the diffractive scintillation.  The observed CP in
pulsars can be several tens of percent, but it may be that some of
this CP results from birefringence in the pulsar magnetosphere itself,
which we do not consider here.  Based on data on RM variations
associated with the Vela pulsar we predict that scintillation-induced CP 
is likely to be observable.  We suggest here that the pulse to pulse    
variation in CP observed in some pulsars is due to this effect.

There is a strong case that the small ($\sim 0.5 - 0.05$\%) degree of CP
exhibited by some compact  extragalactic sources is due to
scintillation-induced CP.  The high RMs (e.g., Udomprasert
\et 1997\markcite{Udom}) and proposed strong magnetic fields (Rees 
1987\markcite{Rees}) near the
cores of AGN are sufficient to generate the observed degree of CP in
association with refractive scintillation.

A prediction is that scintillation-induced CP should reverse sign
randomly, such that the average value is zero. There is some evidence for
this for pulsars, for which the pulse-averaged CP is small
compared with the relatively high CP in individual pulses. However, there is
no strong evidence for reversals in the sense of CP for extragalactic
sources. To determine whether or not this is compatible with the theory
requires further consideration of the nature of RM fluctuations.
One obvious point concerns the distinction between mean RMs and the
variance of the RM. If the length scale over which the magnetic field
changes orientation is short compared to the path length through the
medium, the variance in the difference between the phases of the left-
and right-hand  polarized wavefronts may be very much larger than the
mean value.  This point is exemplified by studies of the ISM which seek
to relate the mean electron density along a line of sight to its
scattering properties.  Based on the variation of $\langle n_e \rangle$
observed in the  Galaxy, $\langle (\delta n_e)^2 \rangle$ is predicted to
vary by 1.3 orders of magnitude, however the observed variation  is
considerably larger, at 4.2 orders of magnitude  (Cordes, Weisberg \&
Borkiakoff 1985\markcite{Cordesetal}).  For the same reason, the mean RM 
along a given line of sight
may not reflect the degree of scintillation-induced CP expected.

The generation of CP due to ultra fine scale structure in the  magnetic
field also needs to be addressed.  If the observed density fluctuations
in the ISM are the result of magnetohydrodynamic turbulence, one expects
similarly fine scale structure in the magnetic field.  This also is
expected to boost the importance of scintillation-induced CP in the ISM.
Further detailed modeling of this effect is warranted.

\acknowledgments
We thank Mark Walker for suggesting the effect of RM gradients and
Ron Ekers for helpful discussions relating to extragalactic sources.

\appendix

\section{Appendix: Evaluation of $g(\beta)$}
We wish to evaluate the integral
\begin{eqnarray}
\langle D(r_{ij}) \rangle &=& \int d^2{\bf x}_i d^2{\bf x}_j  p({\bf x}_i)
p({\bf x}_j) \left(
\frac{\vert {\bf x}_i -{\bf x}_j \vert}{r_{\rm diff} }\right)^{\beta-2}
\nonumber \\
&\null& = \left( \frac{1}{r_{\rm diff}} \right)^{\beta-2} \int dx_i dx_j dy_i
dy_j \, p(x_i)p(x_j) p(y_i) p(y_j) \nonumber \\
&\null& \qquad \qquad \qquad \qquad \times 
\left[(x_i-x_j)^2 + (y_i-y_j)^2 \right]^{(\beta-2)/2}. \label{dijequn}
\end{eqnarray}
Using the expansion
\begin{eqnarray}
(x+y)^\gamma =  \sum_{i=0}^{\infty} \left( \prod_{j=0}^{i-1} (\gamma-j)
\right) x^{\gamma-i} y^i,
\end{eqnarray}
equation (\ref{dijequn}) becomes a series of products of 2-dimensional 
integrals:
\begin{eqnarray}
\langle D(r_{ij}) \rangle &=& \left( \frac{1}{r_{\rm diff}} \right)^{\beta-2}
\sum_{m=0}^{\infty} \left( \prod_{k=0}^{m-1}
\left(\frac{\beta-2}{2}-k\right) \right)
\int dx_i dx_j \, p(x_i) p(x_j) (x_i-x_j)^{\beta-2-2m}
\nonumber \\ &\null& \times
\int dy_i dy_j \, p(y_i) p(y_j) (y_i-y_j)^{2m}.
\end{eqnarray}
An observer sees speckles over the entire scattering disk of radius
$r_{\rm ref}$, and this provides a cutoff to the distribution
$p({\bf x}_i)$. For the purposes of calculation, we approximate the outer
boundary of the speckle pattern as a {\it square} of area $r_{\rm ref}$.
We then have
\begin{eqnarray}
p({\bf x}_i) = \left\{ \begin{array}{ll} 1/r_{\rm ref}^2 & |x_i| <
r_{\rm ref}/2, |y_i| < r_{\rm ref}/2, \\
0 & {\rm otherwise.} \end{array} \right.
\end{eqnarray}
Making the substitutions $u_{x} = x_i - x_j$, $u_y = y_i-y_j$, $v_x =
x_i + x_j$ and $v_y = y_i + y_j$, we then have
\begin{eqnarray}
 \langle D(r_{ij}) \rangle &=& \frac{1}{4 r_{\rm ref}^4}
 \left( \frac{1}{r_{\rm diff}} \right)^{\beta-2} 
 \sum_{m=0}^{\infty} \left( \prod_{k=0}^{m-1}
 \left(\frac{\beta-2}{2}-k\right) \right)
 \int_{A} du_x dv_x \, u_x^{\beta-2-2m} \nonumber \\
 &\null& \qquad \qquad \qquad \qquad \qquad \qquad \qquad 
 \times \int_{A} du_y dv_y \,
 u_y^{2 m},
\end{eqnarray}
where the integration area is taken into account as follows:
\begin{eqnarray}
    \int_A du_x du_y &=&
    \int_{0}^{r_{\rm ref}} du_x \int_{0}^{r_{\rm ref}-u} dv_y +
    \int_{0}^{r_{\rm ref}} du_x \int_{u-r_{\rm ref}}^{0} dv_y 
    \nonumber \\ &\null& +
    \int_{-r_{\rm ref}}^{0} du_x \int_{-r_{\rm ref}-u}^{0} dv_y +
    \int_{-r_{\rm ref}}^{0} du_x \int_{0}^{r_{\rm ref}+u} dv_y.
\end{eqnarray}
Using
\begin{eqnarray}
\int_A du_x dv_x \, u_x^{\beta - 2 - 2 m} =\frac{4 r_{\rm ref}^{\beta -
2m}}{(\beta-2 m)(\beta-2 m -1)},
\end{eqnarray}
provided $\beta-2 m \neq 1$ and $\beta - 2 m \neq 0$, and
\begin{eqnarray}
\int_A du_y dv_y \, u_y^{2 m} =\frac{4 r_{\rm ref}^{2 + 2 m}}{(2 + 2 m)(2 m +1)},
\end{eqnarray}
we obtain
\begin{eqnarray}
\langle D(r_{ij}) \rangle &=&
 4 \left( \frac{r_{\rm ref} }{r_{\rm diff}} \right)^{\beta-2}
  \sum_{m=0}^{\infty} \left( \prod_{k=0}^{m-1}
  \left(\frac{\beta-2}{2}-k\right) \right)
  \left[\frac{1}{(2m + 2)(2m+1)} \right] \nonumber \\
  &\null& \qquad \qquad \qquad \qquad \times \left[ 
  \frac{1}{(\beta-2m)(\beta-2m-1)} \right] \\
  &\equiv&  g(\beta) \left( \frac{r_{\rm ref} }{r_{\rm diff}}
\right)^{\beta-2}. \label{gdefn}
\end{eqnarray}

\section{Derivation of ray-bending due to a Faraday wedge}
Here we derive the result of the ray-bending due to the Faraday wedge
model discussed in \S\ref{FaradayWedge}.
The relative displacement of the right- and left-polarized wavefronts
is derived using Hamiltonian equations for a ray. These are
\begin{eqnarray}
{d{\bf x}\over dt}={\partial\omega_\sigma({\bf k},{\bf x},t)
\over\partial{\bf k}},
\qquad
{d{\bf k}\over dt}=-{\partial\omega_\sigma({\bf k},{\bf x},t)
\over\partial{\bf x}},
\qquad
{d\omega\over dt}={\partial\omega_\sigma({\bf k},{\bf x},t)
\over\partial t}. \label{HamiltonEqns}
\end{eqnarray}
In a stationary medium, assumed here, the third equation is not
relevant. In the paraxial approximation one separates $\bf x$ into
$z$, $\bf r$, with the $z$ axis along ${\bf k}=k{\bkappa}$.
The relation $k=n \omega/c$ is used after combining the first and
second of  equations (\ref{HamiltonEqns}) to yield an expression for the
angular deviation of the ray in terms of the refractive index
variations on the scattering screen
\begin{eqnarray}
{d(n_\sigma\,{\bkappa})\over dz}=
{\partial n_\sigma\over\partial{\bf r}},
\end{eqnarray}
where the arguments of $n_\sigma(\omega,{\bg\kappa},z,{\bf r})$
are suppressed.  The change, $\delta{\bkappa}_\sigma$, across a
wedge of thickness $\Delta z$ is given by
\begin{eqnarray}
\delta{\bkappa}_\sigma={1\over n_\sigma}
\int_0^{\Delta z}dz\,
{\partial n_\sigma\over\partial{\bf r}}.
\end{eqnarray}
In the weak anisotropy limit one writes the refractive index in
terms of polarization dependent and
independent terms $n_\sigma=n+\sigma\Delta n/2$ with
$\Delta n=n_+ - n_-$ to simplify this to
\begin{eqnarray}
\delta {\bkappa}_{\sigma} =
\frac{1}{n} \int_{0}^{\Delta z} dz\, \left\{
\left[1 - \frac{\sigma \Delta n}{2 n} \right]
{\partial n\over\partial{\bf r}}
+ \frac{\sigma}{2} {\partial \Delta n\over\partial{\bf r}} \right\},
\label{deltakappa}
\end{eqnarray}
neglecting terms of order $(\Delta n)^{2}$.
Thus, if the screen is a distance $L$ from the observer, the
relative displacement of the centers of the images in the two
modes on the observer's plane is $L\,\Delta{\bzeta}$, as given by 
equation (\ref{deltazeta1}).

% \begin{eqnarray}
% \Delta {\bzeta} = -{\Delta n\over n} \frac{c}{\omega}
% \int_0^{\Delta z}dz\,
% {\partial n\over\partial {\bf r}}
% +\frac{c}{\omega} \int_0^{\Delta z}dz\,
% {\partial \Delta n\over\partial {\bf r}}.
% \end{eqnarray}
% In terms of the magnetoionic parameters, assuming both $X=(\nu_p/\nu)^2
% \ll1$ and $Y=\nu_B / \nu \ll1$, one has $n=1-\half X$, $\Delta n=XY \cos
% \theta$. Then one finds that the dominant contribution to
% $\Delta\,\zeta$ is due to the gradient of
% $\int_0^{\Delta z} (c/\omega) XY \cos \theta \, dz$,
% which we write as $\phi_{\rm RM}= \lambda^2 \,
% {\rm RM}$.  The displacement between the
% left- and right-circularly polarized
% images on the observer's screen is
% \begin{eqnarray}
% {\bf \Delta x} = \frac{L}{k} \nabla_{\perp} \phi_{\rm RM},
% \end{eqnarray}
% which is the result quoted in \S\ref{FaradayWedge}.

\references

\reference{Backer78}
Backer, D.C.,
1978,
\apj,
222,
L9

\reference
{BW}
Born, M. \& Wolf, E., Principles of Optics, Oxford: Permagon Press, 
1965

% \reference{Car97a}
% Carilli, C.L., Perley, R.A, R\"ottgering, H.J.A \& Miley, G.K.,
% {1997},
% in IAU Symposium 175: Extragalactic Radio Sources, eds. C. Fanti and R.
% Ekers (Kluwer, Dordrecht), p. 159.

\reference
{Caw}
Cawthorne, T.V., Wardle, J.F.C, Roberts, D.H., Gabuzda, D.C. \&
Brown, L.F.,
{1993},
\apj,
{416},
496

\reference
{Cordesetal}
Cordes, J.M., Weisberg, J.M. \& Boriakoff, V.,
{1985},
\apj,
{288},
221

% \reference{Dreher}
% Dreher, J.W., Carilli, C.L. \& Perley, R.A.,
% {1987},
% \apj,
% {316},
% 611

\reference{Goodman85}
Goodman, J.W., Statistical Optics, New York: Wiley, 1985

\reference
{GoodNar}
Goodman, J. \& Narayan, R.,
{1989},
\mnras,
{238},
995

\reference
{Gwinnetal}
Gwinn, C.R., Ojeda, M.J., Britton, M.C., Reynolds, J.E., Jauncey, D.L.,
King, E.A., McCulloch, P.M., Lovell, J.E.J., Flanagan, C.S., Smits, D.P.,
Preston, R.A., Jones, D.L.,
{1997}
\apj,
{483},
L53

\reference{Gwinnetal98}
Gwinn, C.R., Britton, M.C., Reynolds, J.E., Jauncey, D.L., King, E.A.,
McCulloch, P.M., Lovell, J.E.J. \& Preston, R.A. 1998, \apj, {505}, 928

\reference{Hamiltonetal}
Hamilton, P.A., Hall, P.J., Costa, M.E.,
{1985)},
\mnras,
{214},
5

\reference{Han98}
Han, J.L., Manchester, R.N., Xu, R.X. \& Qiao, G.J.,
{1998},
\mnras,
300,
373

\reference
{Johnstonetal}
Johnston, S., Nicastro, L. \& Koribalski, B.,
{1998},
\mnras,
{297},
108

\reference
{Johnston1259}
Johnston, S., Manchester, R.N., Lyne, A.G., D'Amico, N., Bailes, M.,
Gaensler, B.M., Nicastro, L.,
{1996},
\mnras,
{279},
1026

\reference
{Jones77a}
Jones, T.W., O'Dell, S.L.,
{1977a},
\apj,
{214},
552.

\reference
{Jones77b}
Jones, T.W., O'Dell, S.L.,
{1977b},
\apj,
{215},
236

\reference
{Kazbegi}
Kazbegi, A.Z., Machabeli, G.Z. \& Kelikidze, G.I. 1991, \mnras,
253, 377

\reference
{Kom}
Komesaroff, M. M., Roberts, J. A., Milne, D. K., Rayner, P. T. \&
Cooke, D. J.,
{1984},
\mnras,
{208},
409

\reference
{LeggWest}
Legg, M.P.C., Westfold, K.C.,
{1968},
\apj,
{154},
499

\reference{MMcp1}
Macquart, J.-P., Melrose, D.B. 2000 (Paper I), Phys. Rev. E. (submitted)

\reference
{Man75}
Manchester, R.N., Taylor, J.H. \& Huguenin, G.R., 1975, \apj, 196, 83

\reference
{Matveenko}
Matveenko, L.I., Pauliny-Toth, I.I.K., Baath, L.B., Graham, D.A.,
Sherwood, W.A. \& Kus, A.J.,
{1996},
\aap,
{312},
738

\reference
{Melrose93a}
Melrose, D.B. 1993a, J.\ Plasma Phys.,
{50}, 267

\reference
{Melrose93b}
Melrose, D.B. 1993b, J.\ Plasma Phys.,
{50}, 283

\reference
{MelMac}
Melrose, D.B., Macquart, J.-P.,
{1998},
\apj,
{505},
921

\reference{MinSpa}
Minter, A.H., Spangler, S.R.,
{1996},
\apj, {458}, 194.

\reference
{Narayan}
Narayan, R.,
{1992},
Phil Trans R Soc Lond A,
{341},
151.

\reference
{Pachol}
Pacholczyk, A.G.,
{1973},
\mnras,
{163},
29P.

\reference
{dePater} de Pater, I., Weiler, K.W.,
{1982},
\mnras,
{198},
747

\reference
{radhakrishnan}
Radhakrishnan, V. 1992, in T.H.~Hankins, J.M.~Rankin \& J.A.~Gil (eds),
The magnetospheric structure and emission mechanisms of radio pulsars,
proceedings of IAU 128, Pedagogical University Press, Zielona G\'ora,
p.\ 367.

\reference
{RadRank90}
Radharkrishnan, V., \& Rankin, J.M., {1990}, \apj, {352}, 258.

\reference
{Rankin}
Rankin, J.M.,  Campbell, D.B., Isaacman, R.B., Payne, R.R.,
{1988},
\aap,
{202},
166

\reference
{Rees}
Rees, M.J.,
{1987},
\mnras,
{228},
47P.

\reference
{DHRob}
Roberts, D.H., Kollgaard, R.I., Brown, L.F., Gabuzda, D.C. \& Wardle,
J.F.C.,
{1990},
\apj,
{360},
408

\reference
{JARob}
Roberts, J.A., Cooke, D.J., Murray, J.D., Cooper, B.F.C, Roger, R.S.,
Ribes, J.-C., Biraud, F.,
{1975}
AuJPh,
{28},
325.

\reference{Saikia}
Saikia, D.J., \& Salter, C.J., {1988}, \araa, {26}. 93.

\reference
{Sazonov}
Sazonov, V.N.,
{1969},
Sov. Phys. JETP
{29},
578

\reference
{Taylor}
Taylor, G.B.,
{1998},
\apj,
{506},
637

% \reference{Tayloretal}
% Taylor, G.B., Ge, J.-P. \& Barton, E.,
% {1994},
% \aj,
% {107},
% 1942.

\reference
{Udom}
Udomprasert, P.S., Taylor, G.B., Pearson, T.J. \& Roberts, D.H.,
{1997},
\apj,
{483},
L9

\reference{Walker}
Walker, M.A.,
{1998},
\mnras,
{294},
307

\reference
{Weiler}
Weiler, K. W, de Pater, I.,
{1983},
\apjs,
{52},
293

\reference{zhao93}
Zhao, J.-H. \& Goss, W.M., 1993, in: Sub-arcsecond Radio Astronomy
(eds: R.J.~Davis \& R.S.~Booth), Cambridge University Press,
Cambridge, England, 38

\endreferences

\newpage

\plotone{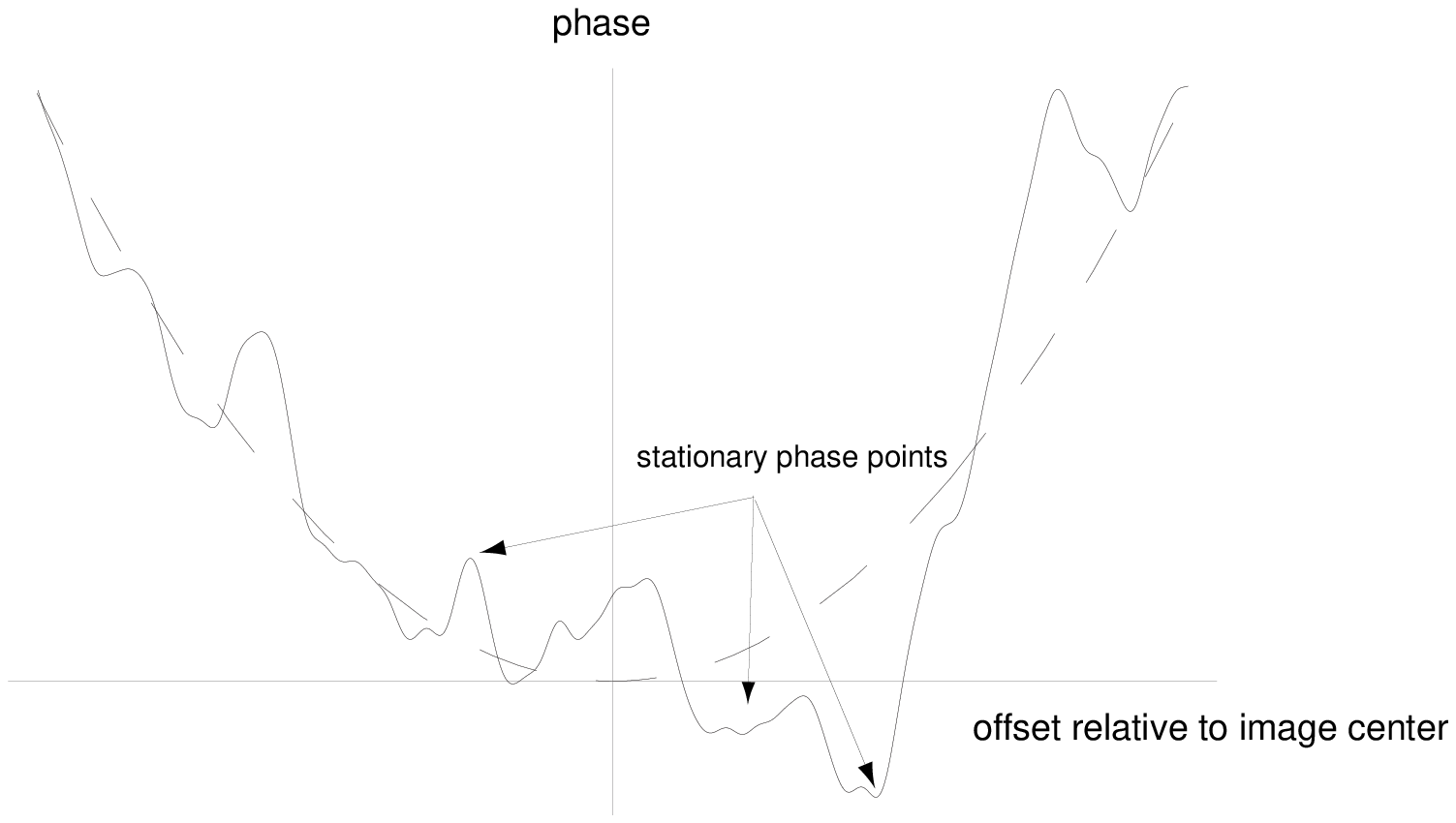}
\figcaption[StatPhase.eps]{An illustration of the scattered wavefront, 
showing the contribution of the geometric phase (dotted), and the 
combined contribution of the scattering medium and geometric phase 
(solid line).  Several points of stationary phase are indicated on 
the diagram.}
%\special{psfile=StatPhase.eps hscale=100 vscale=100}

\newpage

\plotone{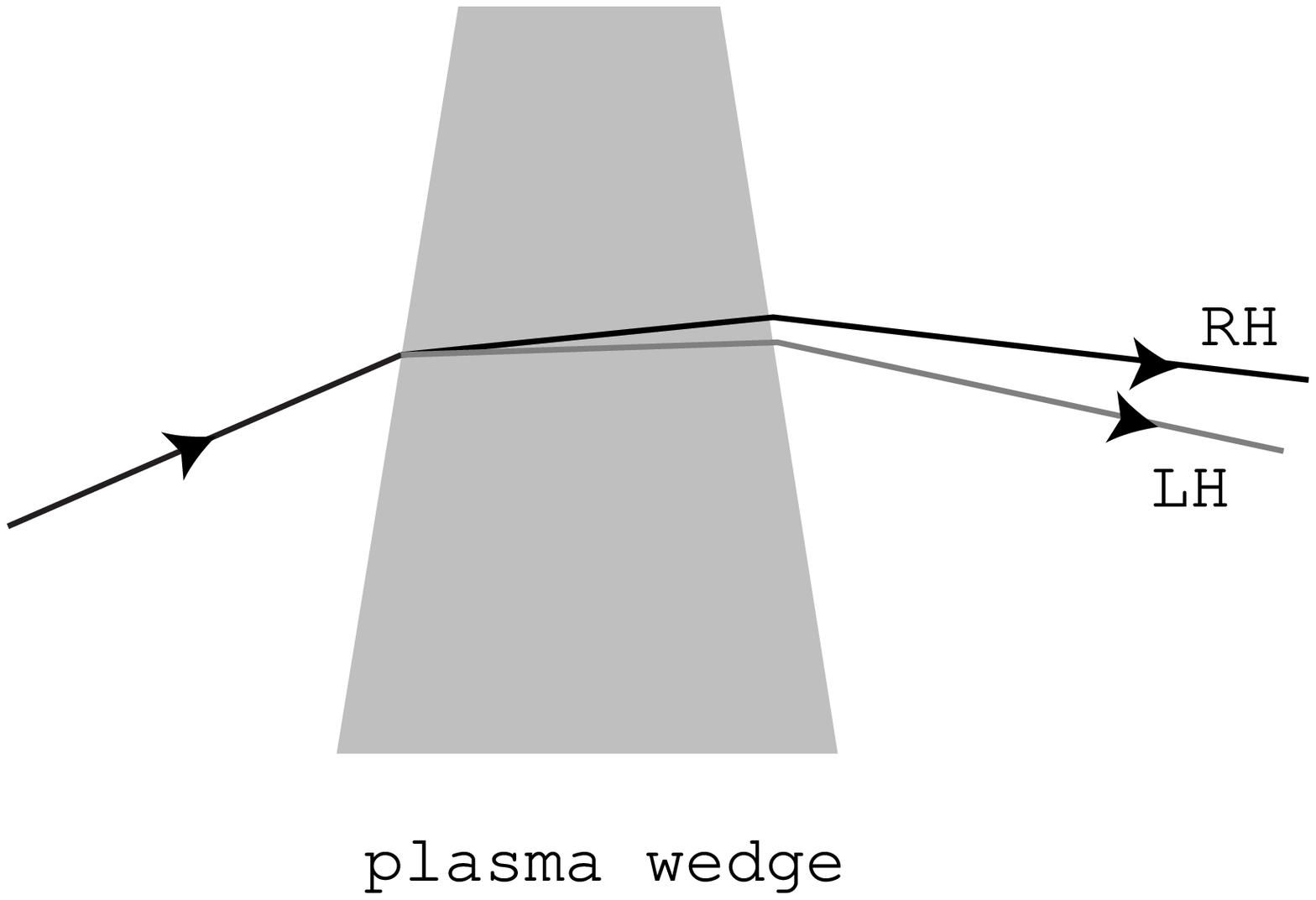}
\figcaption[Fwedge.eps]{A schematic of refraction of rays at a Faraday
wedge: an incident ray splits into two circularly polarized rays
propagating in slightly different directions.}
% \special{psfile=Fwedge.eps hscale=100 vscale=100}

\newpage

\plotone{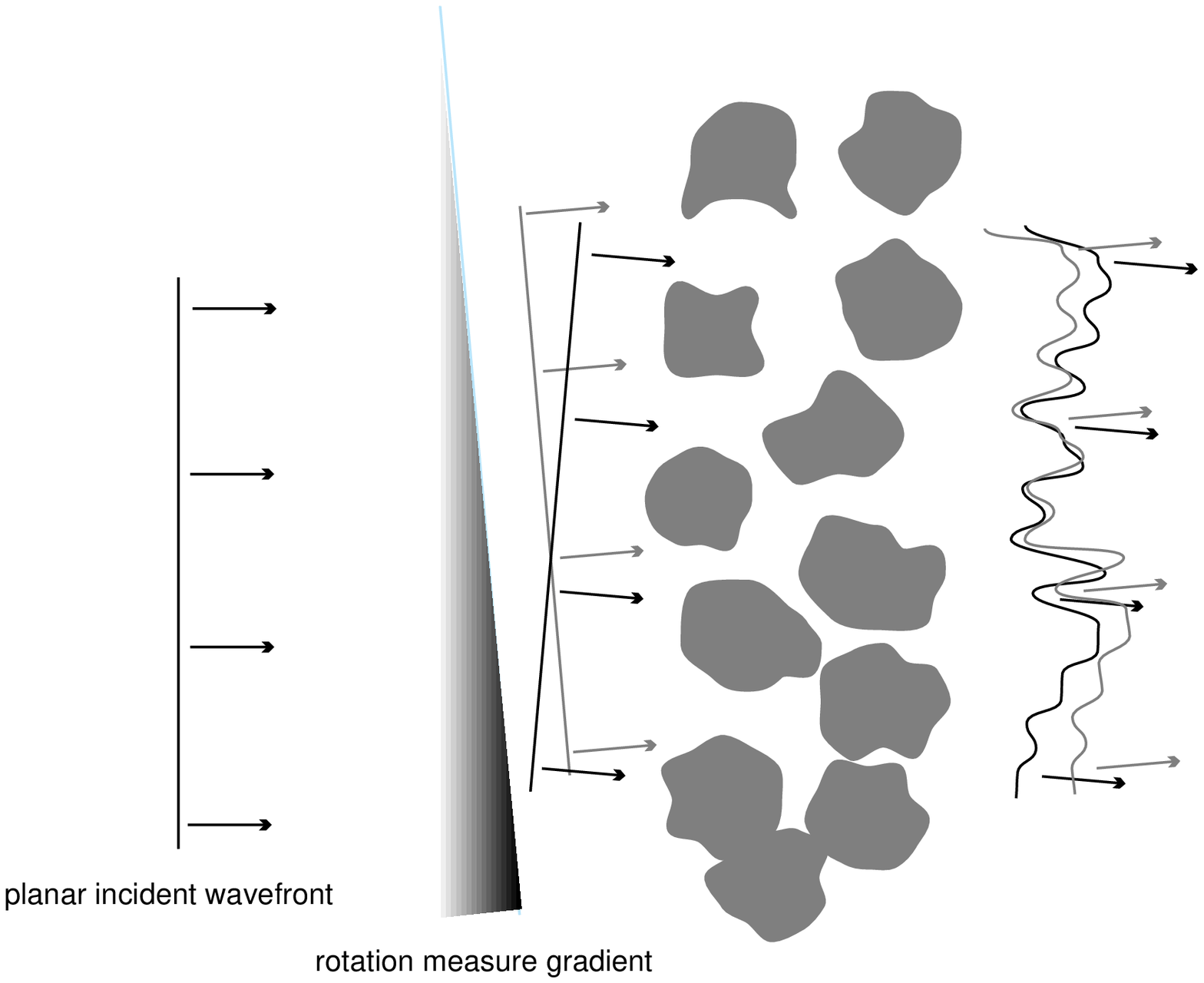}
\figcaption[CRMwedge.eps]{A schematic of the Faraday wedge from the
viewpoint of physical optics.  A  RM gradient causes a
difference in the ray paths of  the left- and right-hand circularly
polarized wavefronts.  Upon  arrival at the Earth, the scintillation
pattern of one wavefront is  slightly displaced with respect to the
other, leading to variability  in the CP.}
% \special{psfile=CRMwedge.eps hscale=100 vscale=100}

\end{document}